\documentclass[]{jingdong}

\microtypesetup{expansion=false}

\usepackage[utf8]{inputenc}
\usepackage[T1]{fontenc}

\usepackage{amsfonts}
\usepackage{amsmath}
\usepackage{amssymb}
\usepackage{booktabs}
\usepackage{xcolor}
\usepackage{enumitem}
\usepackage{url}
\usepackage{xspace}
\usepackage{wrapfig}
\usepackage{array}
\usepackage{listings}
\tcbuselibrary{skins,raster,breakable}

\lstdefinestyle{jncode}{
  language=Python,
  basicstyle=\ttfamily\scriptsize,
  keywordstyle=\color{blue},
  commentstyle=\color{black!50},
  stringstyle=\color{red},
  numbers=left,
  numberstyle=\tiny\color{gray},
  numbersep=4pt,
  firstnumber=1,
  xleftmargin=2.2em,
  framexleftmargin=1.6em,
  frame=single,
  frameround=tttt,
  framerule=0.8pt,
  rulecolor=\color{black!55},
  backgroundcolor=\color{gray!5},
  breaklines=true,
  breakatwhitespace=true,
  columns=flexible,
  keepspaces=true,
  showstringspaces=false,
  tabsize=2,
  escapeinside={(*@}{@*)},
  morekeywords={None,running,wait},
}

\newcolumntype{L}[1]{>{\raggedright\arraybackslash}p{#1}}
\newcolumntype{C}[1]{>{\centering\arraybackslash}p{#1}}

\newcommand{\name}{JoyNexus}
\crefformat{section}{\S#2#1#3}
\Crefformat{section}{\S#2#1#3}
\crefmultiformat{section}{\S#2#1#3}{ and \S#2#1#3}{, \S#2#1#3}{ and \S#2#1#3}
\Crefmultiformat{section}{\S#2#1#3}{ and \S#2#1#3}{, \S#2#1#3}{ and \S#2#1#3}
\crefrangeformat{section}{\S#3#1#4 to \S#5#2#6}
\Crefrangeformat{section}{\S#3#1#4 to \S#5#2#6}

\title{JoyNexus: Service-Oriented Multi-Tenant Post-Training for VLA Models}

\author[1,2,*]{Haoran Sun}
\author[1,3,*]{Wentao Zhang}
\author[1,4]{Junyang Hua}
\author[1,2]{Hedan Yang}
\author[5]{\\ Yongjian Guo}
\author[1,3]{Yifei Zhang}
\author[1,3]{Xiaolong Xiang}
\author[1]{Mingxi Luo}
\author[1]{\\ Jing Long}
\author[1]{Chen Zhao}
\author[1]{Chen Zhou}
\author[1]{Wanting Xu}
\author[1]{Qiming Yang}
\author[1]{\\ Hui Zhang}
\author[1]{Song Wang}
\author[1]{Xiaodong Bai}
\author[1]{Shuai Di}
\author[2]{Xu Chu}
\author[2]{Xiaotie Deng}
\author[1]{Yicheng Gong}
\author[1,\dagger]{Junwu Xiong}

\affiliation[1]{JDT AI Infra}
\affiliation[2]{Peking University}
\affiliation[3]{Beihang University}
\affiliation[4]{Beijing Institute of Technology}
\affiliation[5]{Tsinghua University}

\contribution[*]{Equal Contribution}
\contribution[\dagger]{Corresponding author}

\abstract{
The post-training of Vision-Language-Action (VLA) models is essential due to the diversity of simulators, robot embodiments, and task objectives. Existing compute services, whether offered as direct accelerator rental or batch-workload submission, typically allocate an exclusive set of GPU and CPU resources to a single tenant. While this paradigm maximizes client flexibility, it burdens users with infrastructure adaptation, and the fixed card-hour accounting model renders short or bursty workloads both expensive for tenants and inefficient for the service provider. To address these challenges, we present \textbf{\name{}}, a unified service for multi-tenant VLA supervised fine-tuning, reinforcement learning, and evaluation. \name{} decouples the Training Model Service, Inference Model Service, and Environment Service, each accessed through APIs and backed by resident shared base models with tenant-specific slots. Tenants can directly invoke high-level semantic APIs for training, rollout, and evaluation, or compose custom algorithms using lower-level APIs and their assigned endpoints. Multiple tenants submit workloads concurrently; their action modules, optimizers, rollout records, and policy versions remain isolated, and the service is scheduled by the global Training Queue and Inference Queue. To further improve multi-tenant training efficiency, \name{} introduces group batching for heterogeneous VLA data schemas that share a compatible model-facing prefix, enabling a single shared backbone forward pass over grouped samples. Finally, we evaluate \name{} through workload simulation and a group-batching pipeline in a realistic embodied scenario. Results show that, compared with isolated single-tenant execution, \name{} reduces aggregate GPU time and improves service utilization via cross-tenant scheduling on shared resources.

}

\date{\today}
\correspondence{Junwu Xiong (\texttt{xiongjunwu.1@jd.com})}

\begin{document}
\maketitle

\section{Introduction}
\label{sec:introduction}

Vision-language-action (VLA) models offer a promising pathway toward integrating the multimodal comprehension capabilities of foundation models with embodied agents and systems.
Models such as RT-2~\citep{zitkovich2023rt}, OpenVLA~\citep{kim2024openvla}, $\pi_0$~\citep{black2024pi0}, and GR00T~\citep{bjorck2025gr00t} demonstrate that pretrained vision-language representations can be effectively adapted to robotic manipulation and control tasks. 
While large-scale pretraining endows these models with general capabilities, post-training remains necessary due to the diversity of data schemas and robot embodiments across different tasks. 
Moreover, tasks of varying difficulty demand different post-training paradigms. 
A typical development workflow may involve supervised fine-tuning (SFT) on labeled data~\citep{kim2024openvla,kim2025fine,octo}, evaluation on simulators~\citep{gu2023maniskill2,liu2023libero,mees2022calvin,james2020rlbench}, online and offline reinforcement learning (RL) with simulators~\citep{li2025simplevla,vlarft,vlac,sun2026rl}, and data pipeline orchestration~\citep{ramos2021rlds,openxembodiment,cadene2024lerobot}, among other stages.

For cloud service providers, existing paradigms either rent dedicated compute resources to tenants~\citep{armbrust2009above} or execute user-submitted workloads on shared clusters~\citep{verma2015large,lab2026mint}.
Both approaches grant tenants full control over execution but require them to manage complex infrastructure dependencies, which is particularly challenging for VLA training involving heterogeneous model and simulator environments.
Moreover, as VLA models are often moderate in scale, tenant-designed distributed training can lead to poor accelerator utilization.
Therefore, fixed card-hour pricing is inefficient for small, iterative, and bursty VLA workloads, where GPUs may remain idle during rollout, data loading, evaluation, or environment synchronization.

Tinker-style systems have shown that model post-training can be exposed through programmatic APIs while abstracting away much of the distributed execution substrate~\citep{tml2026tinker}. 
More recently, agentic RL systems have explored rollout-as-a-service and training-service separation~\citep{zhang2026prorl,zhu2026opentinker}, and managed LLM training infrastructures provide analogous abstractions for training and serving at scale~\citep{lab2026mint,relax2026}. 
The fundamental innovation of these works lies in freeing tenants from low-level infrastructure concerns so they can focus on algorithm development. 
By exposing composable service APIs from which tenants construct their own workloads and pipelines, providers can schedule shared resources at finer granularity.
Nevertheless, current frameworks mainly target language-model post-training, while VLA-specific RL and evaluation remain underexplored.

In this paper, we present \textbf{\name}, a framework that reframes VLA post-training as a multi-tenant training service. 
Inspired by the Tinker-style service-oriented paradigm, \name{} separates Training and Inference Model Services from the Environment Service, which abstracts both interactive environments and offline datasets.
Users submit workload specifications describing the model, data or environment, training mode, evaluation target, and resource requirements. 
The platform provisions tenant-specific runtime objects and routes requests to shared services.
Users can invoke high-level operations, including training, rollout, evaluation, export, and adapter synchronization, or compose custom algorithms through lower-level service APIs.

\name{} is built on three key insights.
First, RL, SFT, and evaluation for VLA share common infrastructure, including model inference, environment interaction, and data exchange, motivating a unified decomposition into Training, Inference, and Environment Services.
Second, parameter-efficient VLA post-training typically freezes the shared VLM backbone while updating tenant-specific action modules, enabling lightweight multi-tenant adaptation.
Third, the service-oriented design supports concurrent tenant scheduling and improves GPU utilization through group batching across heterogeneous workloads.

Our contributions are summarized as follows:
\begin{itemize}
    \item We propose a client-server service abstraction for embodied post-training that separates tenant-private computation from shared infrastructure, maintaining workload isolation while the server controls resource allocation, session placement, and service routing.
    \item We design a three-component backend—the Training Model Service, Inference Model Service, and Environment Service—that jointly supports SFT, RL, and evaluation, with base models kept memory-resident and tenant-specific modules mounted in isolated slots.
    \item Experimental results show that, compared to classic single-tenant serial workload processing, \name{} achieves higher efficiency through improved resource utilization.
\end{itemize}

The rest of the paper is organized as follows. 
We discuss related systems and service frameworks in Section~\ref{sec:related} and introduce the preliminary modeling of VLAs and workflows in Section~\ref{sec:preliminaries}. 
Section~\ref{sec:architecture} presents the \name{} architecture, including the overall structure, unified SFT/RL/evaluation workflows, and multi-tenant scheduling. 
Section~\ref{sec:experiments} reports implementation details and the evaluation of \name{} on efficiency improvements compared to native serving. 
Finally, Section~\ref{sec:conclusion} concludes with directions for future work.

\section{Related Work}
\label{sec:related}

In this section, we discuss related literature, including service-oriented post-training systems, distributed training frameworks for foundation models, and recent VLA post-training and evaluation methods.

\subsection{Tinker-style Services for Foundation Models}

Tinker-style systems expose model post-training through programmatic APIs while hiding much of the distributed execution substrate from users. Tinker lets users express fine-tuning logic against service primitives rather than manually managing worker placement, distributed model initialization, and artifact movement \citep{tml2026tinker,knight2025tinker}. The key idea is not remote execution, but a separation of concerns: users compose learning programs, while the service owns resource allocation, model residency, scheduling, synchronization, and persistence.

Recent systems extend this service view to language-model post-training. OpenTinker~\citep{zhu2026opentinker} studies concern separation in agentic reinforcement learning and builds on the veRL~\citep{sheng2024hybridflow} line of RLHF infrastructure. ProRL Agent~\citep{zhang2026prorl} exposes rollout-as-a-service for multi-turn agent RL, and MinT~\citep{lab2026mint} studies managed infrastructure for training and serving large numbers of LLMs. AReaL decouples rollout generation from policy training to improve asynchronous RL execution, while MARLaaS extends this direction to multi-tenant RL-as-a-service through a shared base model, tenant-specific LoRA adapters, and independently scheduled rollout, environment, and training stages~\citep{fu2025areal,yu2026marlaas}. The Twinkle system used in our implementation follows the same engineering philosophy~\citep{modelscope2026twinkle}. These systems primarily target language models or text-based agents; VLA post-training additionally requires simulator sessions, heterogeneous action schemas, rollout records, evaluation protocols, adapter manifests, and policy-version synchronization. To our knowledge, \name{} is the first Tinker-style system designed for a complete VLA-specific post-training ecosystem, including SFT, RL, rollout, evaluation, parameter export, and inference synchronization.

\subsection{Distributed Training Frameworks for Foundation Models}

Distributed training frameworks provide the computational substrate on which a service such as \name{} can run. Megatron-LM introduced practical tensor model parallelism for training multi-billion-parameter transformers \citep{shoeybi2019megatron}. ZeRO and DeepSpeed reduce optimizer, gradient, and parameter redundancy to improve memory efficiency at very large model scales \citep{rajbhandari2020zero}. Colossal-AI provides a unified interface for combining data, tensor, pipeline, sequence, and heterogeneous parallelism \citep{li2021colossal}. Ray provides a general distributed execution substrate for task and actor workloads, and RLlib builds distributed RL abstractions on top of it \citep{moritz2018ray,rllib}. These systems focus on scaling computation and memory for one or more training workloads; \name{} addresses a complementary layer above them, namely how tenant workloads are admitted, routed, isolated, and connected to resident model and environment services.

Large-model post-training frameworks are closer to \name{} in workflow structure. DeepSpeed-Chat and OpenRLHF package RLHF training pipelines for chat-style language models, while HybridFlow models RLHF as a dataflow with hierarchical APIs for efficient orchestration~\citep{yao2023deepspeed,openrlhf,sheng2024hybridflow}. RLinf separates logical RL workflows from physical execution planning, RLinf-VLA specializes this abstraction for VLA reinforcement learning, and RL-VLA$^3$ shows that asynchronous simulator, generator, and trainer groups can accelerate VLA RL~\citep{yu2025rlinf,zang2025rlinf,sun2026rl}.

Multi-tenant training and serving systems provide complementary mechanisms. LobRA jointly fine-tunes multiple LoRA adapters over heterogeneous tenant data while sharing the base model~\citep{lin2025lobra}. At the serving layer, Clipper introduced adaptive batching for low-latency prediction, AlpaServe studies statistical multiplexing across multiple models, and vLLM and Sarathi improve the memory and scheduling efficiency of large-model inference~\citep{clipper,alpaserve,vllm,sarathi}. Punica and S-LoRA batch inference across LoRA adapters that share one resident base, while dLoRA dynamically orchestrates both requests and adapters across serving replicas~\citep{punica,slora,wu2024dlora}. \name{} can use such systems as backend implementations, but addresses a complementary layer in which client-server service boundaries, dual training and inference scheduling paths, tenant-private state, artifact visibility, policy routing, and VLA environment interaction are first-class objects.

\subsection{SFT, RL, and Evaluation for Vision-Language-Action Models}

VLA foundation models motivate the workloads addressed by \name{}. RT-1 demonstrated transformer-based real-world robotic control at scale, and RT-2 showed that web-scale vision-language knowledge can transfer to robotic action~\citep{rt1,zitkovich2023rt}. PaLM-E extended multimodal language modeling to embodied reasoning, and Open X-Embodiment collected cross-robot datasets and RT-X models for transferable robot learning~\citep{palme,openxembodiment}. OpenVLA made a large open-source VLA model available for manipulation and demonstrated downstream fine-tuning~\citep{kim2024openvla}. Octo studied a generalist robot policy trained on diverse robot datasets, while $\pi_0$ and GR00T-style policies further illustrate the diversity of action-generation mechanisms and robot embodiments~\citep{octo,black2024pi0,bjorck2025gr00t}. HPT uses a shareable policy trunk with embodiment-specific interfaces, and X-VLA uses lightweight embodiment-specific soft prompts to learn from heterogeneous robot platforms~\citep{wang2024hpt,zheng2026xvla}. These models make post-training infrastructure important because each user may bring different datasets, simulators, action horizons, robot states, or action-module layouts.

SFT, RL, and evaluation are all central to VLA development. OpenVLA, OpenVLA-OFT, and Octo study downstream fine-tuning and adaptation, while Open X-Embodiment, RLDS, and LeRobot provide reusable data and tooling for robot learning pipelines~\citep{kim2024openvla,kim2025fine,octo,openxembodiment,ramos2021rlds,cadene2024lerobot}. SimpleVLA-RL, VLAC, and VLA-RFT show that RL or reinforcement fine-tuning introduces online loops among policy inference, simulator interaction, rollout storage, and model updates~\citep{li2025simplevla,vlac,vlarft,sun2026rl}. Evaluation is commonly measured through simulator and benchmark suites such as ManiSkill2, LIBERO, CALVIN, and RLBench~\citep{gu2023maniskill2,liu2023libero,mees2022calvin,james2020rlbench}. \name{} does not propose a new VLA model, objective, or benchmark; it provides a service substrate that can run these SFT, RL, rollout, and evaluation workloads for many tenants while sharing resident base models and preserving tenant-specific artifacts.

\section{Preliminaries}
\label{sec:preliminaries}

This section introduces the VLA model and workflow abstractions used in the rest of the paper.

\subsection{Typical VLA Model Structure}
\label{sec:prelim-vla-model}

Current vision-language-action models typically convert multimodal input context into an action sequence through a shared perception-language backbone and an embodiment-specific action module. The model first encodes visual observations, language instructions, robot state, and optional history information into a latent representation. The action-producing module then maps the latent representation to robot actions, which can be a diffusion or flow-matching action expert~\citep{chi2025diffusion}, or even a lightweight MLP or projection head~\citep{kim2025fine}. Representative models include OpenVLA~\citep{kim2024openvla}, the $\pi$ series~\citep{black2024pi0}, and the GR00T series~\citep{bjorck2025gr00t}.

\begin{figure}[!htbp]
    \centering
    \includegraphics[width=0.98\linewidth]{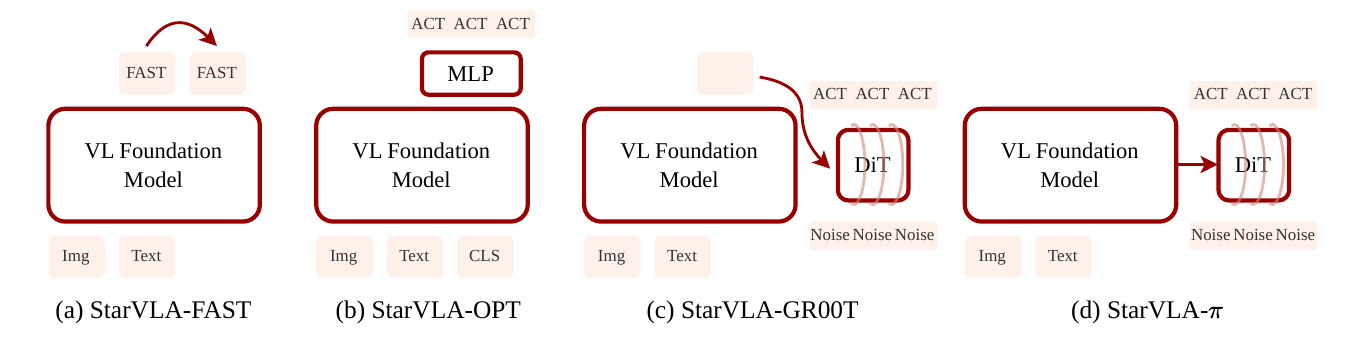}
    \caption{Adapted from Figure 2 of StarVLA~\citep{starvla2026lego}. Representative VLA model structures: recent VLA models commonly connect a pretrained vision-language backbone to action-specific decoding modules.}
    \label{fig:prelim-vla-model}
\end{figure}

More recently, StarVLA~\citep{starvla2026lego} proposes a Lego-style decomposition between the vision-language base model and action-specific heads, as illustrated in Figure~\ref{fig:prelim-vla-model}. This facilitates flexible composition between different VLMs and sophisticated action expert designs. We also observe that VLA post-training often keeps the base VLM fixed, as the task-specific data is relatively narrow and may degrade the general capability of vision-language comprehension. This motivates the design principle of \name{}: the base model or shared prefix is expensive and often reusable across tenants, whereas the action module, optimizer, and processors are typically tenant-specific. In the rest of the paper, we use \emph{resident base model} to refer to the reusable shared component maintained by the service. We use \emph{action module} as an umbrella term for an action head, action expert, lightweight adapter, or policy suffix, and \emph{tenant-private action state} for the module and associated metadata belonging to one tenant workload.

\subsection{SFT, RL, and Evaluation Workflows}
\label{sec:prelim-workflows}

\begin{figure}[!htbp]
    \centering
    \includegraphics[width=0.96\linewidth]{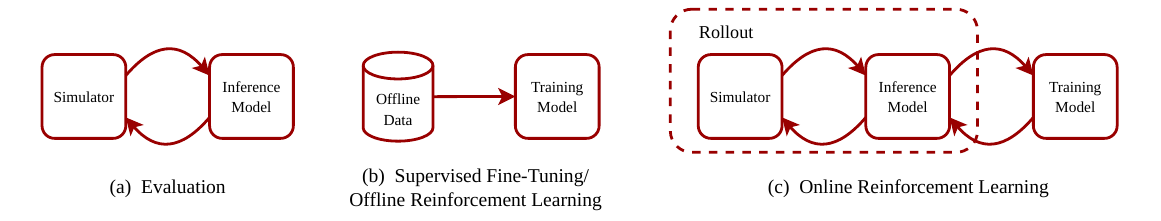}
    \caption{Representative VLA post-training workflows. Evaluation loads a fixed policy and records metrics; SFT and offline RL consume offline data and update model parameters; online RL closes the loop among simulator interaction, inference, rollout storage, training, and parameter synchronization.}
    \label{fig:prelim-workflows}
\end{figure}

VLA post-training commonly alternates among supervised fine-tuning, reinforcement learning, and evaluation. As abstracted in Figure~\ref{fig:prelim-workflows}, these workflows differ in where data comes from and whether model parameters are updated, but they reuse many of the same system components: data access, model inference, training updates, parameter export, and artifact storage. This shared structure is the main workflow-level abstraction used by \name{}.

\paragraph{Supervised fine-tuning.}
In SFT, a tenant provides an offline dataset of demonstration records. Each record typically contains observations, a language instruction, an optional robot state, and target actions. The training process samples supervised batches from the dataset, runs the VLA model, computes an action prediction loss, and updates the tenant's trainable modules. After a fixed number of training steps, the system exports a parameter artifact, such as an adapter checkpoint, that can later be loaded for inference or evaluation.

\paragraph{Reinforcement learning.}
In RL, the data source is an online or simulated environment rather than a fixed demonstration dataset. The current policy receives observations, predicts actions through the inference path, and sends those actions back to the environment. The resulting transitions, rewards, termination signals, and policy metadata are written as rollout records. A training job then consumes those rollout records, performs an RL update, exports the updated tenant parameters, and synchronizes the serving policy used by future rollouts. Compared with SFT, RL therefore adds a closed loop among inference, environment interaction, rollout storage, training, and parameter synchronization.

\paragraph{Evaluation.}
Evaluation measures a fixed policy without changing its parameters. An evaluation workload loads a chosen adapter or policy revision, binds an environment or dataset, runs inference, and records metrics. This can be considered as part of the rollout process, and can therefore reuse the same Inference Model Service, Environment Service, and artifact store as training.
\section{\name{} Service Architecture}
\label{sec:architecture}

This section presents the service architecture of \name{}.
We first describe its client--server structure and the division of responsibilities between the control plane and the execution services.
We then explain how the same backend primitives compose SFT, RL, and evaluation workflows.
Finally, we introduce a scheduling strategy for tenants that share a frozen base model, with an emphasis on group batching when individual requests contain only small batches.

\subsection{Overall Client-Server Structure}
\label{sec:overall-structure}

\begin{figure}[t]
    \centering
    \includegraphics[width=0.96\linewidth]{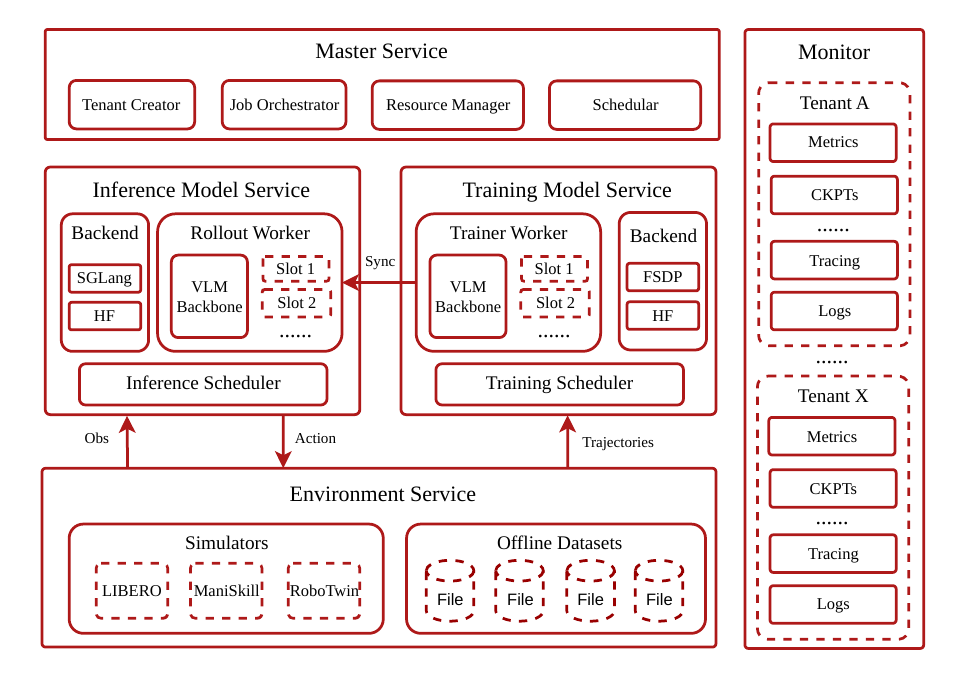}
    \caption{\name{} service architecture. The Master Service compiles user intent into tenant-scoped workloads and coordinates the resident Training Model, Inference Model, and Environment Services. The Training Queue and Inference Queue decouple optimization data flow from latency-sensitive prediction requests, while the monitoring plane exposes workload and queue progress.}
    \label{fig:jdtinker-runtime}
\end{figure}

As shown in Figure~\ref{fig:jdtinker-runtime}, \name{} consists of three logical parts: user-facing workload specifications, a Master Service that forms the control plane, and resident model and environment services that form the execution plane.
A user declares the base model, task type, trainable parameterization, environment or dataset, and scheduling requirements for an RL, SFT, evaluation, or custom workload.
The Master Service validates user intent and translates it into a tenant-scoped workload, whereas the execution services maintain the expensive shared runtime and process the resulting requests.
This separation keeps workflow semantics at the client and control-plane level while allowing the server to manage placement, concurrency, data transfer, and state.

\subsubsection{Master Service}
\label{sec:tenant-specific-layer}

The \textbf{Master Service} integrates tenant creation, workload orchestration, resource management, and scheduling.
Its tenant creator strictly validates a workload specification against backend capabilities, derives a deterministic schema signature for each tenant, and compiles the specification into a backend configuration.
The workload orchestrator materializes the required service roles and lifecycle: RL workloads connect rollout, inference, environment, and training; SFT workloads connect an offline-data producer to training; and evaluation workloads reuse inference and environment services without creating an optimization path.
Consequently, user intent is separated from backend-specific process layout and communication details.

Resource management and scheduling operate on explicit service, workload, and training-job state.
The controller places long-lived services in accelerator placement groups and monitors their health, while tenant-specific action modules, optimizer states, policy versions, environment sessions, and checkpoints remain separately identified.
The control plane exposes the Training Queue and Inference Queue, which receive work issued by different tenant workloads and dispatch it according to configurable priorities.
Their scheduling policies are described in Section~\ref{sec:group-batching}.
This division provides state isolation and predictable execution without requiring the base model to be replicated for every tenant.

\subsubsection{Model and Environment Services}
\label{sec:tenant-agnostic-backend}

The execution plane contains the resident services reused across workloads that share a base model.
Following the Lego-like VLA design of StarVLA~\citep{starvla2026lego}, the model services separate a shared vision-language base model from trainable action modules.
In the current design, these tenant-specific modules primarily instantiate action experts for different robot embodiments and action spaces.
The \textbf{Training Model Service} keeps the shared base model resident and maintains tenant-specific action modules and training states, including optimizer state and policy version. We refer to its update worker as the \emph{actor}.
A training job selected by the Training Scheduler activates the corresponding tenant state, performs the update, and exports only that tenant's parameter payload; the shared base model is neither duplicated nor included in the payload.
The \textbf{Inference Model Service} maintains tenant-indexed serving states and synchronizes updated action-module parameters from training.

The \textbf{Environment Service} exposes session-based \texttt{reset}, \texttt{step}, and \texttt{close} operations for online interaction (\texttt{batch\_step} is an optional throughput optimization).
Different tenants may be routed to different external simulator services, while an SFT tenant bypasses environment interaction and reads demonstrations through a dedicated data producer.
During RL or evaluation, environment observations generate requests for the Inference Queue and the resulting actions are returned to their originating sessions.
The Training Queue connects rollout and SFT producers to the training consumer, and the checkpoint path stores tenant parameter payloads, optimizer states, and a manifest that identifies each tenant and version.
Together, these mechanisms keep model, data, and environment lifecycles decoupled while preserving tenant identity across service boundaries.

These three execution services are decoupled and accessible through specific APIs, which is intentionally organized two-layered.
The lower layer contains the core service primitives in Figure~\ref{lst:service-apis}, together with scheduler and parameter-synchronization operations.
They expose direct control over environment sessions, Training Queue partitions, Inference Queue requests, actor updates, checkpoints, and serving-weight updates, allowing advanced users to construct customized workflows.
The upper layer accepts semantic task declarations---RL, SFT, or evaluation---and lets the Master Service compile them into a sequence of lower-level operations.
The next subsection describes these compositions, with pseudo code deferred to Appendix~\ref{app:api-composition}.

\begin{figure}[t]
\centering
\begin{lstlisting}[style=jncode]
# Training Model Service
train_job({job_id, tenant_id, partition_id, task_type})  # lease; microbatch updates
train_batch({tenant_id, actions, advantages, fwd})       # one optimizer step on a slot
export_params({tenant_id, schema, version})              # tenant payload (no shared base)
save({step, tenant_ids, include_optim})                  # checkpoint params / optimizer

# Inference Model Service
predict({observations, tenant_id, policy_version})       # action (and value) inference
rollout({config, job_id, num_envs, horizon, version})    # env loop composed over predict
load_weights({tenant_id, version, named_tensors})        # sync tenant action-module weights

# Environment Service / Training Queue
reset({session_id, rollout_id}) / step({session_id, actions}) / close({session_id})
queue_put({partition_id, data}) / queue_get({partition_id, batch_size, fields})
queue_clear({partition_id})                              # release a consumed partition
\end{lstlisting}
\caption{Core APIs of the Training Model, Inference Model, and Environment/Queue services in the current implementation (simplified names). The Training Queue data plane is grouped with environment-session operations; latency-sensitive prediction enters the Inference Queue via the Inference Model Service.}
\label{lst:service-apis}
\end{figure}

\subsubsection{System Functionality}
\label{sec:system-functionality}

Beyond model execution and environment interaction, \name{} provides essential runtime capabilities for long-running multi-tenant post-training workloads. 
We highlight three key functionalities: centralized monitoring, fault isolation, and elastic scaling.

\paragraph{Monitoring.}

\name{} adopts a centralized monitoring architecture that separates metric collection from individual workers. 
Instead of embedding tracking logic into each training or rollout component, runtime services asynchronously emit metric events to a dedicated Metrics Service, such as ClearML~\cite{clearml} or WandB~\cite{wandb}.

These events contain three categories of information. 
Control signals, including job states, queue depth, in-flight requests, and policy staleness, are consumed by the Master Service for admission control and resource management. 
Learning signals, such as training losses, rollout returns, evaluation scores, and policy versions, are forwarded to experiment tracking systems for user-facing analysis. 
System signals, including service health, execution latency, weight synchronization overhead, and accelerator utilization, enable operators to diagnose performance bottlenecks. Meanwhile, checkpoints and workload manifests are stored independently in each tenant's output directory, ensuring that experiment artifacts remain reproducible and auditable even after temporary runtime states are released.

\paragraph{Fault isolation and partial restart.}

To support reliable long-running workloads, \name{} introduces service-level fault isolation through a dedicated Health Manager. 
The Health Manager continuously monitors resident services through heartbeats and explicit error reports.

When a localized failure occurs, the controller performs an in-place restart instead of restarting the entire workload. 
Specifically, only the failed role is terminated and redeployed within the existing resource allocation, with its execution state restored from the recorded progress before resuming service. 
Other components and tenant workloads continue execution without interruption.
A global restart is triggered only when failures affect shared coordination components or exceed the local recovery budget. 
This design confines failures to the smallest possible scope and improves system robustness in multi-tenant environments.

\paragraph{Elastic scaling.}

By decoupling inference and training into independent services, \name{} enables dynamic scaling of rollout capacity without interrupting the optimization process.

Scaling operations are performed asynchronously. 
During scale-out, new rollout engines are launched, synchronized with the current tenant policy weights, and gradually admitted into the serving pool. 
During scale-in, engines first drain outstanding requests before releasing allocated resources. 
Concurrent scaling operations are serialized to maintain serving consistency and enable safe rollback under partial failures.

This elasticity is particularly important for multi-tenant VLA training, where environment interaction workloads can vary significantly across tasks. 
\name{} can dynamically allocate additional rollout resources during inference-intensive phases and reclaim idle accelerators after workloads complete, without modifying tenant-side training logic.

\subsection{Server Backend: Unifying Multi-Tenant SFT, RL, and Evaluation}
\label{sec:unified-workflows}

\begin{figure}[t]
    \centering
    \includegraphics[width=0.97\linewidth]{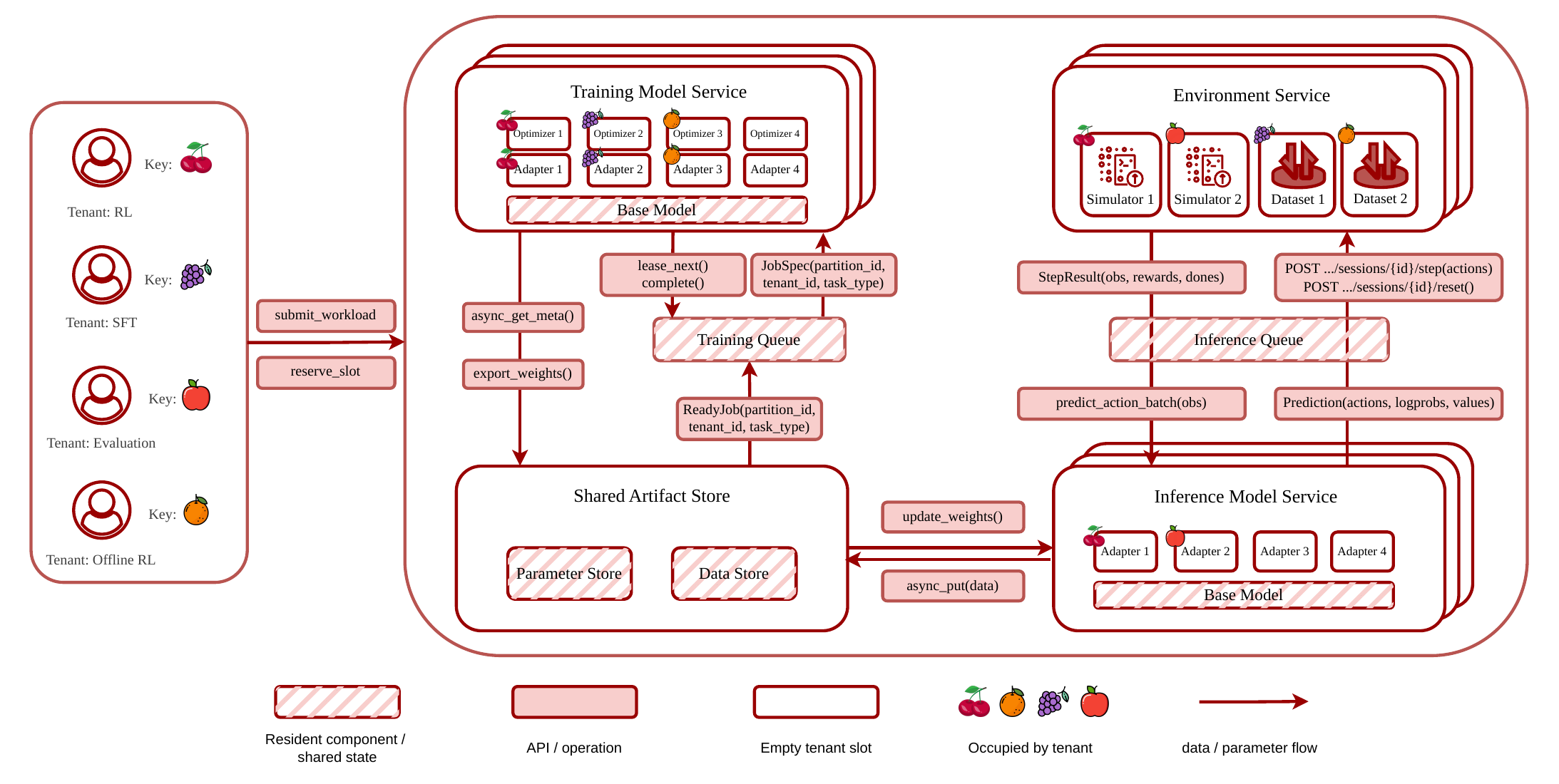}
    \caption{Detailed multi-tenant runtime flow. The Training Scheduler coordinates rollout and demonstration batches consumed by training, whereas the Inference Scheduler coordinates action requests generated by rollout and evaluation. Tenant parameter versions and optimizer states remain independent throughout both queues.}
    \label{fig:jdtinker-detailed}
\end{figure}

\begin{figure}[!htbp]
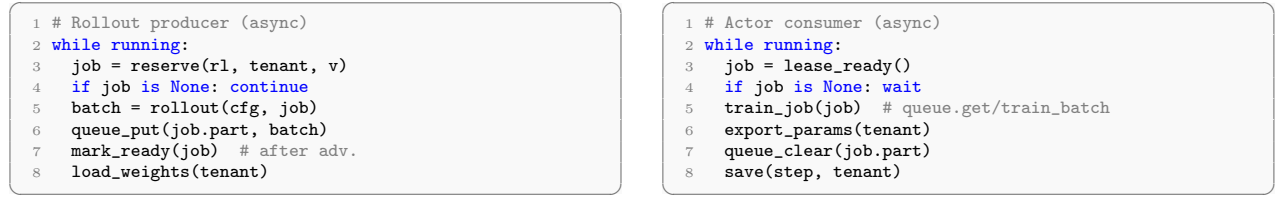

\centering
\begin{minipage}[t]{0.48\linewidth}
\begin{lstlisting}[style=jncode]
# Rollout producer (async)
while running:
  job = reserve(rl, tenant, v)
  if job is None: continue
  batch = rollout(cfg, job)
  queue_put(job.part, batch)
  mark_ready(job)  # after adv.
  load_weights(tenant)
\end{lstlisting}
\end{minipage}
\hfill
\begin{minipage}[t]{0.48\linewidth}
\begin{lstlisting}[style=jncode]
# Actor consumer (async)
while running:
  job = lease_ready()
  if job is None: wait
  train_job(job)  # queue.get/train_batch 
  export_params(tenant)
  queue_clear(job.part)
  save(step, tenant)
\end{lstlisting}
\end{minipage}
\caption{Simplified asynchronous RL logic. Left: rollout loop writing Training Queue partitions. Right: actor loop consuming ready jobs. The loops progress independently and couple only through the queue and scheduler.}
\label{lst:rl-logic}
\end{figure}

Figure~\ref{fig:jdtinker-detailed} illustrates how \name{} composes a common set of backend primitives into complete post-training workloads.
Each training job carries a tenant identifier, task type, schema signature, and, when applicable, a behavior-policy version.
The Training Scheduler manages the training job's control state, and the Training Queue associates its data with a separate partition identifier.
Inference calls generated during rollout and evaluation enter the Inference Queue and retain their originating tenant and environment-session identities.
This dual-queue separation permits the same training consumer to process RL and SFT training jobs while latency-sensitive inference and read-only evaluation follow an independent scheduling path.
Figures~\ref{lst:rl-logic} and~\ref{lst:sft-logic} compose the core APIs from Figure~\ref{lst:service-apis} as two asynchronous loops coupled only through the Training Queue; stage leases and staleness checks are deferred to Appendix~\ref{app:api-composition}.

\paragraph{Reinforcement Learning Closed Loop.}
The RL flow forms a closed loop across environment interaction, inference, and training, but \emph{rollout production and actor consumption run as concurrent loops}.
The rollout side repeatedly reserves a job, executes \texttt{rollout} (environment \texttt{reset}/\texttt{step} plus \texttt{predict}), and \texttt{queue\_put}s the trajectory partition.
Independently, the actor side leases ready jobs, runs \texttt{train\_job}, publishes weights with \texttt{export\_params}/\texttt{load\_weights}, and \texttt{queue\_clear}s the partition.
The two sides communicate only through job state and queue partitions, so collecting the next rollout need not wait for the previous \texttt{train\_job} to finish.

\paragraph{Supervised Fine-tuning Path.}
The SFT path keeps the same asynchronous producer--consumer shape, but the producer samples demonstrations instead of calling \texttt{rollout}/\texttt{predict}.
It \texttt{queue\_put}s a demonstration partition and marks the job ready immediately; the actor loop is unchanged and may interleave SFT jobs with RL jobs on the shared Training Model Service.
No behavior-policy staleness check is required; \texttt{load\_weights} is needed only when the updated tenant must be served for inference.

\begin{figure}[t]
\centering
\begin{minipage}[t]{0.48\linewidth}
\begin{lstlisting}[style=jncode]
# SFT producer (async)
while running:
  job = reserve(sft, tenant)
  if job is None: continue
  demos = sample_demos(cfg)
  queue_put(job.part, demos)
  mark_ready(job)  # immediate
  # no load_weights
\end{lstlisting}
\end{minipage}
\hfill
\begin{minipage}[t]{0.48\linewidth}
\begin{lstlisting}[style=jncode]
# Actor consumer (async)
while running:
  job = lease_ready()
  if job is None: wait
  train_job(job)  # train_batch
  export_params(tenant)
  queue_clear(job.part)
  save(step, tenant)  # optional
\end{lstlisting}
\end{minipage}
\caption{Simplified asynchronous SFT logic. The producer writes demonstration partitions without environment interaction; the actor loop reuses the same \texttt{train\_job} contract as in Figure~\ref{lst:rl-logic}.}
\label{lst:sft-logic}
\end{figure}

\paragraph{Evaluation Path.}
Evaluation is read-only with respect to model parameters.
An evaluation declaration names one or more trainable tenants as targets; at the configured interval, the rollout service selects a target, resolves its current cached parameter version once, and records the resulting metrics.
This path submits requests to the Inference Queue and reuses the evaluation environment, but it does not create a Training Queue partition, invoke the actor optimizer, or advance the target tenant's policy version.
Resolving the version at invocation time is important when evaluation overlaps with training, because a benchmark should not silently observe several policy revisions within one run.

\subsection{Multi-Tenant Group Batching}
\label{sec:group-batching}

The central scheduling challenge in \name{} is to coordinate training and inference requests issued concurrently by different tenants.

\paragraph{Training requests.}
The Training Queue represents optimization work at training-job granularity.
An RL training job typically contains all trajectories collected in one rollout, whereas an SFT training job contains a batch of demonstrations produced by its dataloader; the same abstraction also accommodates other offline-data workloads, such as offline RL.
By default, the Training Scheduler leases ready training jobs in FIFO order.
When priorities are configured, it selects the highest-priority ready training job while preserving FIFO order within each priority class.
The Training Model Service then activates the action module and optimizer state identified by the training job's tenant ID and consumes its samples as consecutive microbatches of the configured size.

Offline workloads can obtain samples directly from a resident dataset, whereas online RL must wait for environment interaction.
Without admission control, readily available offline data could therefore occupy the Training Queue faster than online trajectories are produced.
To prevent offline workloads from crowding out online training jobs, \name{} limits the number of concurrently admitted training jobs for each tenant--task policy.
These in-flight limits bound the associated number of queued samples even when an entire offline dataset is immediately available.

\paragraph{Inference requests.}
The Inference Queue receives observation batches generated during rollout and evaluation, as well as standalone prediction requests submitted through the model API.
Unlike training jobs, an inference request may contain only a small batch because an environment exposes limited parallelism or because a user requests an individual prediction.
Executing such requests strictly one by one would lead to fluctuating load and low accelerator utilization.
To address this, \name{} applies group batching across inference requests: compatible requests are accumulated for the shared base-model forward, and the output features are then partitioned to their respective action modules.
The detailed mechanism and scheduling policy are described below.

Dynamic batching across different requests is a key technique to improve throughput and resource utilization in online systems.  
The basic strategy of dynamically adding inference requests to the current batch for autoregressive LLM serving is already widely used by online systems such as SGLang~\citep{zheng2024sglang} and vLLM~\citep{vllm}, as well as by online RL training engines~\citep{slime_github}.  
Directly applying dynamic batching to \name{} poses two challenges:  
First, unlike the unified token modality of language, different inference or training requests may consist of different data schemas and output structures.  
Second, VLA models often perform one-token inference rather than long-sequence generation, as they require environment feedback for each action (chunk).  
Therefore, we design a pipeline for processing requests with different data schemas, combined with a scheduler that monitors the current batch size and waiting time.

\begin{figure}[t]
    \centering
    \includegraphics[width=0.97\linewidth]{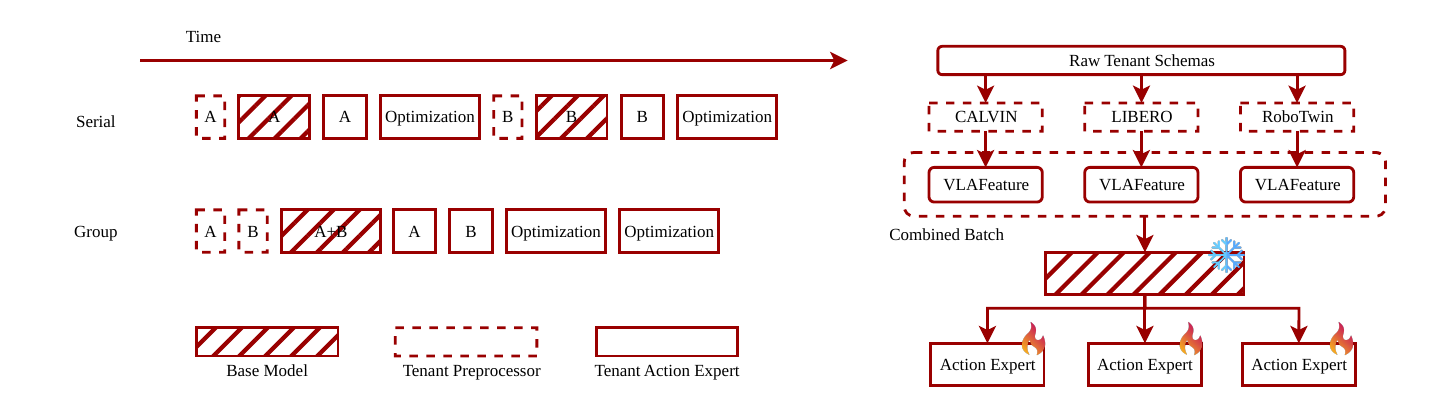}
    \caption{Multi-tenant group batching. Compatible tenant batches are normalized to a shared-prefix representation, processed through a single frozen base-model forward pass, and subsequently partitioned back to tenant-specific action modules, losses, and optimizers.}
    \label{fig:jdtinker-group-batching}
\end{figure}

A demonstration of the processing pipeline is provided in Figure~\ref{fig:jdtinker-group-batching}.  
The left side contrasts serial and group-batched execution: when per-tenant batches are small, grouping increases the physical batch presented to the shared base model.  
The right side shows how the canonical \texttt{VLAFeature} shape is determined for the accepted requests and how the output is subsequently partitioned.  
This strategy amortizes the dominant shared forward pass and improves utilization when compatible requests arrive close together.  
Note that since most VLA models are pretrained on large-scale datasets, they typically adopt a unified data prototype; this naturally constitutes the \texttt{VLAFeature} and is the point at which concatenation occurs.  
The computation output is then distributed to different action experts to complete the subsequent decoding, and gradient computation if needed.

Note that we mainly apply this group batching strategy in the Inference Scheduler rather than the Training Scheduler.  
This is because training jobs usually consist of multiple trajectories that already form reasonably sized micro-batches, thus not requiring further composition across jobs.  
However, the inference service may experience sporadic requests due to the frequent interaction between the inference model and the environment, as well as independent inference API calls.  
The scheduling strategy for the inference service draws from classic queueing theory~\citep{cooper1981queueing}.  
We set a maximum waiting time $T$ and a target batch size $B$.  
The scheduler retrieves samples from requests in a FIFO manner, constructs a batch, and triggers inference when the waiting time reaches $T$ or the current batch size reaches $B$.

\section{Experiments}
\label{sec:experiments}

In this section, we evaluate the resource efficiency and training correctness of \name{}.
Our prototype builds on the service abstractions of existing LLM systems~\citep{relax2026,modelscope2026twinkle} and integrates a local VLA RL stack~\citep{yu2025rlinf,zang2025rlinf,sun2026rl}; the system mechanisms are described in Section~\ref{sec:architecture}, so we focus here on the experimental configs and results.

\subsection{Realistic Multi-Tenant Workload Simulation}
\label{sec:workload-simulation}

We first evaluate \name{} under a realistic mixture of online and offline VLA post-training workloads.
The experiment uses StarVLA~\citep{starvla2026lego} with a Qwen3-VL-4B backbone and four tenants on one 8-GPU node: three RL tenants interact with three simulator workloads---LIBERO and two ManiSkill configurations~\citep{liu2023libero,gu2023maniskill2}---while one SFT tenant continuously submits offline demonstration batches.
We use a 2--2--4 resource layout: two data-parallel GPUs host the Training Model Service actor, two GPUs host the rollout-facing Inference Model Service, and four GPUs run environment simulation.
Within the environment partition, LIBERO and ManiSkill~1 share the first two-GPU pool, while ManiSkill~2 uses the second pool. 
All tenants share the resident base model in both model services but retain tenant-private action heads, value heads, optimizer states, and policy versions.

We use a global actor microbatch of 256 samples, evenly divided across the two actor GPUs (128 samples per GPU), and cap each inference batch at 128 samples.
The scheduler admits RL work in round-robin order and allows at most two outstanding training jobs per RL tenant, while limiting active rollouts to one per tenant and three globally.
Consequently, one rollout from each RL tenant can proceed concurrently, and a tenant can launch its next eligible rollout as soon as capacity becomes available rather than waiting at a cross-tenant barrier.
Ready RL updates take precedence at the actor, whereas the SFT producer prefetches up to 16 batches and uses actor intervals in which the RL tenants are still collecting trajectories.
Once dispatched, an actor update runs to completion; a staleness bound of 128 actor microbatch updates limits how far an in-flight rollout may lag behind its tenant's current policy.

\begin{wraptable}{r}{0.47\linewidth}
    \vspace{-1.0ex}
    \centering
    \small
    \caption{Average utilization of the shared Training and Inference Model Service GPUs. The isolated row is the mean across the three standalone RL traces.}
    \label{tab:vla-service-utilization}
    \begin{tabular}{lrr}
        \toprule
        Execution & Training & Inference \\
        \midrule
        Isolated single-tenant & 20.8\% & 28.1\% \\
        Multi-tenant (\name) & 41.3\% & 37.5\% \\
        Relative improvement & $1.99\times$ & $1.33\times$ \\
        \bottomrule
    \end{tabular}
    \vspace{-1.0ex}
\end{wraptable}

We compare this deployment against an isolated single-tenant baseline constructed from standalone 1-1-2 traces: one GPU for actor training, one for inference, and two for environment simulation.  
The per-GPU actor micro-batch size and simulator density are identical to those of the multi-tenant run.  
For a workload-matched comparison, we select the RL updates whose actor phase begins during the first 90 minutes of the multi-tenant trace and replay the same number of per-tenant updates sequentially in the isolated baseline.  
We also append the same SFT work observed in the matched window.

\begin{figure}[t]
    \centering
    \includegraphics[width=0.98\linewidth]{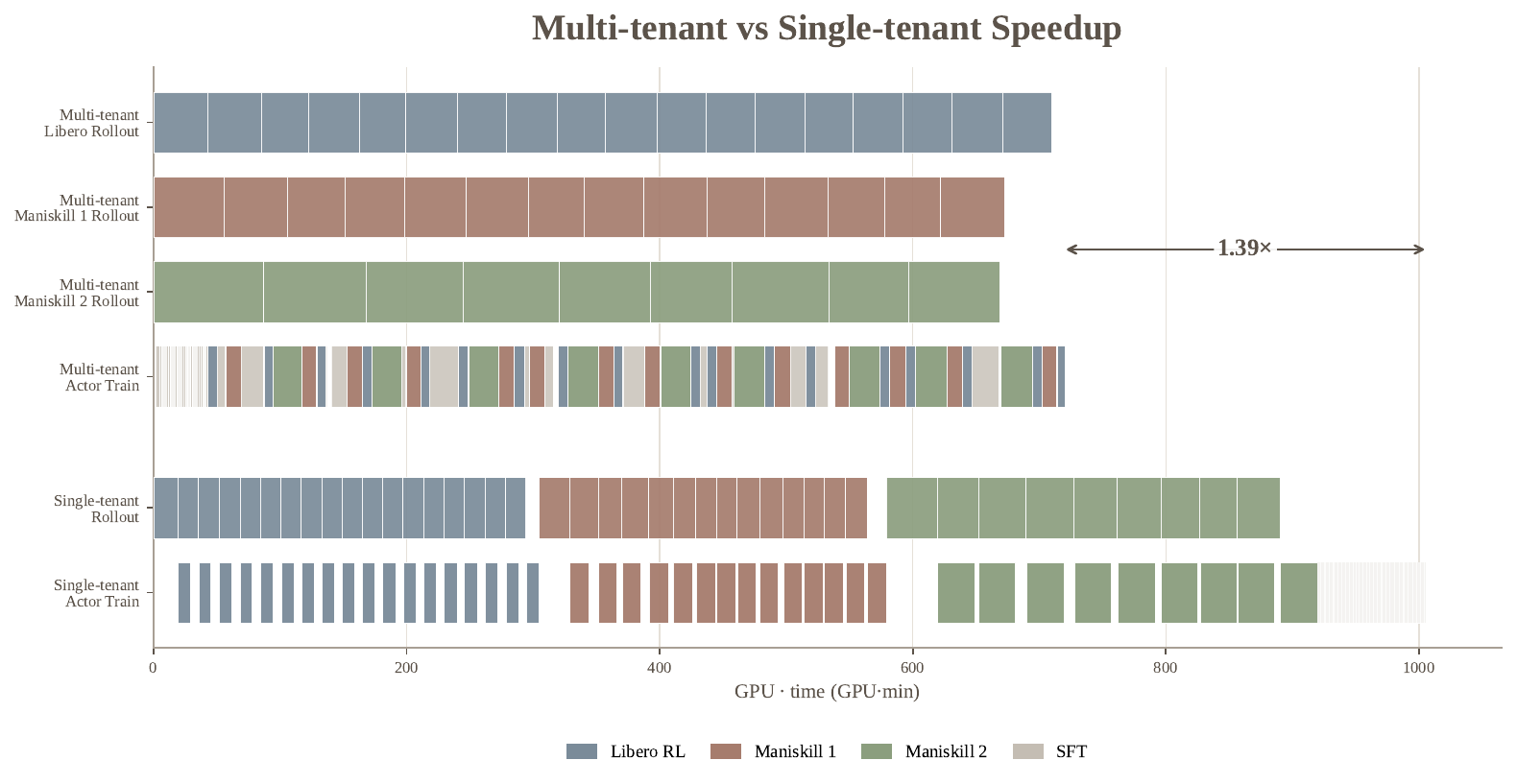}
    \caption{Trace-based comparison of multi-tenant and isolated single-tenant execution for the same VLA workload. The multi-tenant deployment interleaves three RL tenants and one SFT tenant on a shared 8-GPU node, whereas the baseline executes the matched single-tenant workloads sequentially and appends the matched SFT work.}
    \label{fig:vla-multitenant-workload}
\end{figure}

Figure~\ref{fig:vla-multitenant-workload} reports GPU time from the beginning of execution through the completion of the last matched training update.  
The isolated traces expose complementary idle periods: the inference GPU waits during actor updates, while the actor is idle during long simulator interactions.  
In contrast, \name{} overlaps the three asynchronous rollout streams and fills otherwise idle actor intervals with SFT work.  
Although sharing a GPU across different tenants can extend the wall-clock time of an individual tenant's workload, this overlapping reduces aggregate GPU time by 28.3\%, yielding a $1.39\times$ improvement in GPU-time efficiency for the matched workload.  
Table~\ref{tab:vla-service-utilization} corroborates this behavior on the shared model services.  
Cross-tenant scheduling nearly doubles Training Model Service utilization ($1.99\times$) and improves Inference Model Service utilization by $1.33\times$.  
The larger training-side gain reflects both the diversity of RL rollout durations and the offline tenant's ability to make progress while online tenants wait for their environments; the inference service similarly benefits from multiplexing requests arriving at different cadences.

\subsection{Multi-Tenant Group-Batched Execution}
\label{sec:sft-throughput}

In this section, we use a controlled simulation to evaluate group batching under multi-tenant training workloads. Different tenants submit labeled training batches, and each tenant retains its own action-module parameters, optimizer state, loss computation, backward pass, and update. Because group batching only eliminates repeated computation in the resident base model, our efficiency measurements isolate the shared forward stage rather than claiming end-to-end training-throughput improvements. We compare the following two strategies:
\begin{itemize}
    \item \textbf{Serial execution} keeps the base model resident but processes tenant requests one after another, so the shared base-model forward pass is repeated for each tenant.
    \item \textbf{Group execution} canonicalizes each tenant batch and concatenates compatible VLA inputs for one shared base-model forward pass, then splits the resulting features for tenant-private loss computation, backward passes, and updates.
\end{itemize}
We use two representative VLA model families, StarVLA~\citep{starvla2026lego} and OpenPI ($\pi_{0.5}$)~\citep{intelligence2025pi05}.
For StarVLA, we use the model types QwenGR00T and QwenOFT: both employ a Qwen VLM for feature encoding, but the former uses a GR00T-style flow-matching action head for decoding while the latter uses a small linear layer for action decoding.
The $\pi_{0.5}$ model is similar to QwenGR00T in design but has subtle differences in feature extraction and the underlying VLM architecture.
In our experiments, we treat the action head as the tenant-private action module and the VLM as the resident base model.
We use two representative LeRobot-style datasets~\citep{cadene2024lerobot}, LIBERO\footnote{\url{https://modelscope.cn/datasets/lerobot/libero}} and CALVIN\footnote{\url{https://modelscope.cn/datasets/Koorye/calvin-abc-d-lerobot}}, to simulate tenant training batches with different data schemas.

\begin{figure}[t]
    \centering
    \begin{subfigure}[b]{\linewidth}
        \centering
        \includegraphics[width=0.98\linewidth]{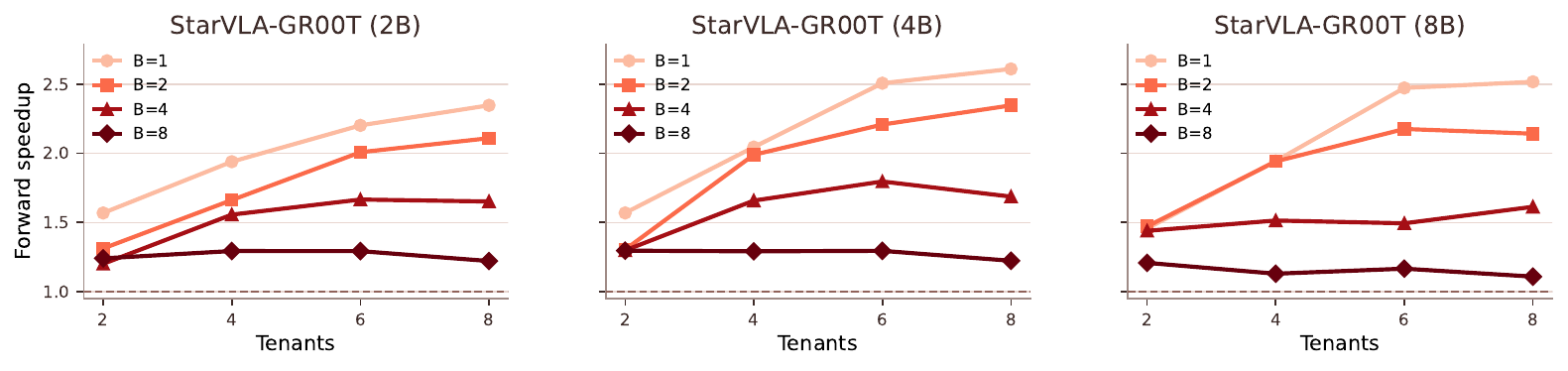}
        \caption{Shared-forward throughput speedup for different QwenGR00T scales in StarVLA.}
        \label{fig:sft-vla-qwengr00t_scale-speedup}
    \end{subfigure}
    \vspace{1ex}
    \begin{subfigure}[b]{\linewidth}
        \centering
        \includegraphics[width=0.98\linewidth]{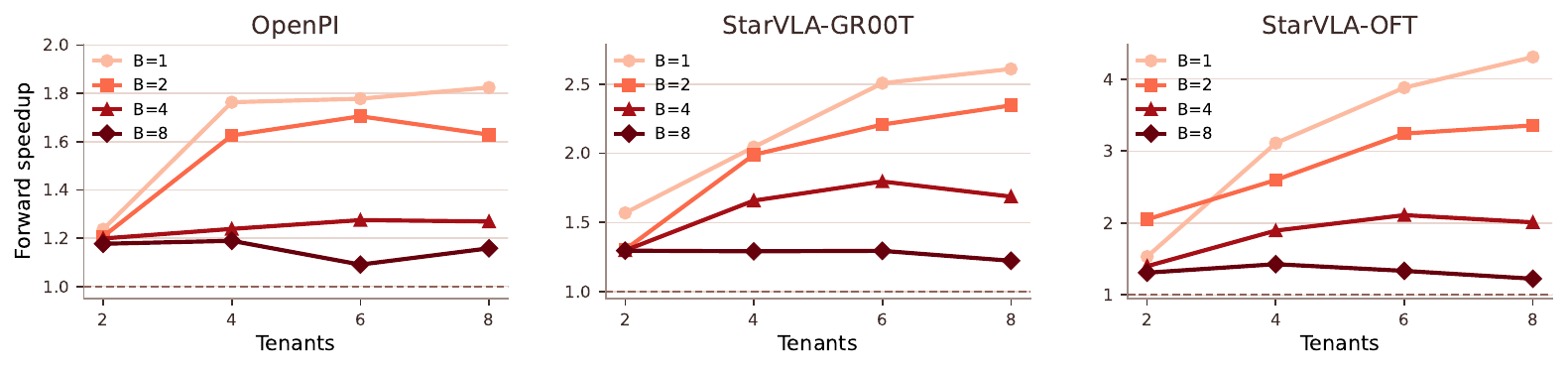}
        \caption{Shared-forward throughput speedup for OpenPI, QwenGR00T, and QwenOFT.}
        \label{fig:sft-vla-openpi_oft_gr00t-speedup}
    \end{subfigure}
    \caption{Shared-forward throughput speedup under mixed-schema multi-tenant training workloads. The metric covers the forward stage that can be shared across tenants, rather than end-to-end training throughput. Group batching is most beneficial when many tenants submit small local batches, matching the intended service regime for bursty tenant updates.}
    \label{fig:combined-speedup}
\end{figure}

We first measure shared-forward throughput for tenant counts $\{2,4,6,8\}$ and local batch sizes $\{1,2,4,8\}$.
Figure~\ref{fig:combined-speedup} reports the speedup of group execution over serial execution for this stage.
In general, group batching becomes more effective as the number of tenants increases and the local batch size decreases.
This trend aligns with the design motivation: when many tenants submit small training batches, each tenant alone underutilizes the resident GPU service, whereas a grouped batch better amortizes base-model execution.
For QwenGR00T models (Figure~\ref{fig:sft-vla-qwengr00t_scale-speedup}), the benefit of group execution grows with VLM scale.
In Figure~\ref{fig:sft-vla-openpi_oft_gr00t-speedup}, OpenPI has a similar scale and architecture to QwenGR00T (4B) and therefore exhibits a similar trend.
For QwenOFT, the tenant-private action module is negligible relative to the base model, so nearly all forward computation occurs in the shared base model and group execution provides a larger improvement.

\begin{figure}[t]
    \centering
    \begin{subfigure}[b]{\linewidth}
        \centering
        \includegraphics[width=0.98\linewidth]{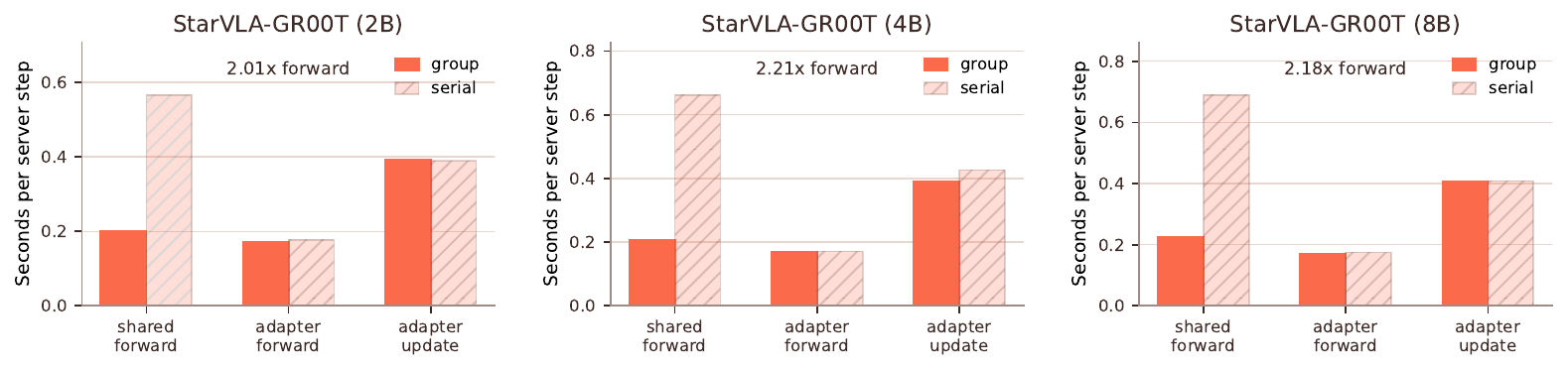}
        \caption{Forward-stage breakdown for different QwenGR00T scales.}
        \label{fig:sft-vla-qwengr00t_scale-breakdown}
    \end{subfigure}
    \vspace{1ex}
    \begin{subfigure}[b]{\linewidth}
        \centering
        \includegraphics[width=0.98\linewidth]{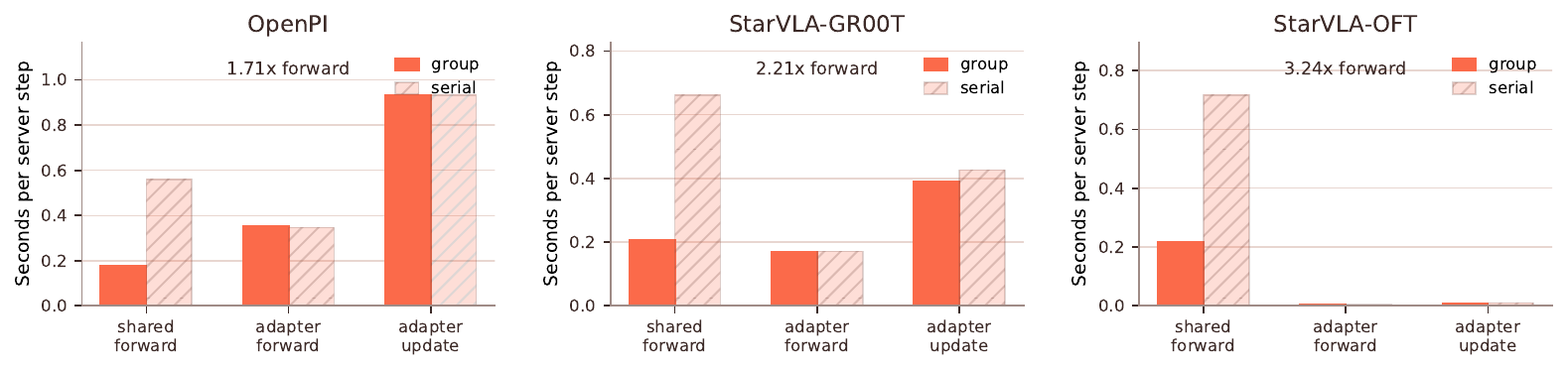}
        \caption{Forward-stage breakdown for OpenPI, QwenGR00T, and QwenOFT.}
        \label{fig:sft-vla-openpi_oft_gr00t-breakdown}
    \end{subfigure}
    \caption{Serial versus group execution for the representative mixed-schema setting with 8 tenants and local batch size 4. Group batching mainly reduces repeated shared-forward work, while tenant-private action-module computation remains separate.}
    \label{fig:combined-breakdown}
\end{figure}

We further break down forward-stage runtime for the setting with eight tenants and a local batch size of 4.
As shown in Figure~\ref{fig:combined-breakdown}, the base-VLM forward time dominates the tenant-specific action-module forward time, even with a large DiT action head.
Group execution substantially reduces repeated VLM forward computation, thereby improving the throughput of the shared-forward stage.
This reduction can also shorten training iterations when the shared forward pass is a major component of iteration time; however, the reported speedup does not include tenant-private backward passes or optimizer updates.
OpenPI shares a similar architecture with QwenGR00T, but its components have different scales: OpenPI's 2B PaliGemma base model is smaller than Qwen-VL-4B, while its action expert contains more parameters than QwenGR00T.

\begin{figure}[t]
    \centering
    \includegraphics[width=0.98\linewidth]{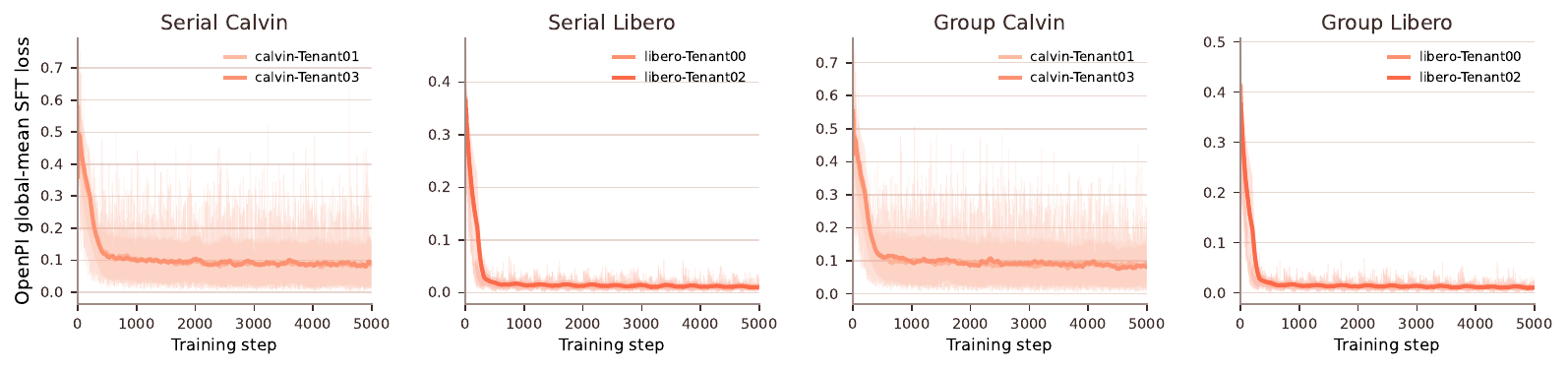}
    \caption{Per-tenant training-loss trajectories under group and serial execution. The experiment uses four OpenPI ($\pi_{0.5}$) tenants with local batch size 2; two tenants train on LIBERO and two on CALVIN, each with separate flow-matching action-head parameters and optimizer state.}
    \label{fig:sft-vla-openpi-loss}
\end{figure}

Finally, we verify that grouping preserves separate tenant-specific training state and does not disrupt per-tenant optimization behavior.
We conduct an experiment with four OpenPI tenants: two train on LIBERO streams and two on CALVIN streams.
Figure~\ref{fig:sft-vla-openpi-loss} compares each tenant's loss trajectory under group execution with that under the resident serial baseline.
All four tenants exhibit the expected early loss descent followed by stable training, and the grouped trajectories closely track their serial counterparts.
Together with the use of separate action-head parameters and optimizer states by construction, these results provide empirical evidence that shared-forward execution preserves per-tenant training-curve behavior. Here, isolation refers to separation of tenant-specific model and optimization state, rather than security or performance isolation.

\section{Conclusion}
\label{sec:conclusion}

This paper presented \name, a multi-tenant service architecture for VLA post-training. \name{} turns SFT, RL, rollout, and evaluation into tenant-private workloads over resident shared model and environment services, so users can express learning intent while the provider manages sessions, routes, action-module state, artifacts, and policy revisions. The current prototype grounds this abstraction with RLinf-based client workflows, Master Service-managed tenant state, the Training Queue and Inference Queue, resident model services, rollout and parameter artifacts, and unified service APIs. The mixed-workload experiment shows that event-driven multi-tenant execution reduces aggregate GPU time while improving both training and inference utilization. The group-batching experiments further show that compatible tenants can share a frozen backbone forward while keeping action modules, losses, optimizers, and checkpoints separate.

\paragraph{Dynamic Resource Adjustment.}
While \name{} supports dynamic addition and removal of model services, tenant workloads remain largely tied to fixed routes and resource assignments during execution. 
A more adaptive design could leverage runtime signals to dynamically adjust service routing and resource allocation without changing user-visible algorithm semantics. 
Such scheduling policies can respond to queue pressure, simulator latency, and tenant priority, improving utilization and reducing waiting time.

\paragraph{Pricing and Productization.}
Beyond system efficiency, \name{} requires appropriate pricing and priority mechanisms for practical deployment. 
A key challenge is designing charging strategies that balance tenant cost, platform utilization, and service guarantees. 
Future work will explore usage-based and priority-aware pricing schemes that account for resource consumption, workload characteristics, and latency requirements.

\paragraph{User Flexibility and Isolation. }
\name{} aims to let users focus on VLA algorithm design rather than infrastructure assembly, but VLA workloads require several migration paths. Some users may run a simulator provided by the platform, others may upload a Docker environment, and others may expose an external environment API. The algorithm layer raises a similar question: users may need to override data processing, rewards, rollout logic, or training code. More flexibility increases adoption but also introduces security, correctness, and efficiency risks. Future versions should define safe extension points, sandboxing rules, attack detection, and performance contracts for user-provided code and environments. For heterogeneous VLA SFT, the same principle applies to data schemas: canonicalization can enable sharing, but only when masks, action dimensions, camera slots, and state conventions preserve the semantics required by each tenant.

\paragraph{Service-Aware Algorithm Design. }
The service substrate also opens algorithmic opportunities. With tenant permission and privacy safeguards, the platform may be able to detect related datasets, tasks, or action schemas across users and use them for data augmentation, transfer, or better initialization. The current Training Queue and Inference Queue intentionally use straightforward FIFO-like semantics with compatibility checks. Richer schedulers could form dynamic training groups based on observed schemas, batch sizes, deadlines, and resource pressure, improving both utilization and learning throughput. Overall, we view \name{} as a valuable step toward programmable, multi-tenant VLA post-training services.

\bibliographystyle{assets/plainnat}
\bibliography{reference}

\FloatBarrier
\appendix
\section{Composition of VLA Workloads}
\label{app:api-composition}

This appendix provides an implementation-oriented expansion of the high-level workflows in Section~\ref{sec:unified-workflows}.
The pseudocode follows the current control flow, but omits Ray remote-call syntax, distributed collectives, and model-specific tensor operations.
These omissions make explicit the service composition without tying the description to a particular deployment size.

\subsection{Dual-Queue Scheduling Abstraction}

At the architecture level, \name{} exposes two independent scheduling paths, summarized in Table~\ref{tab:app-dual-queue}.
The Training Queue represents optimization work whose payload must remain available across rollout generation, preprocessing, and actor consumption.
The Inference Queue represents latency-sensitive prediction work whose result is returned directly to a rollout or evaluation session.
This distinction is semantic: a logical queue denotes the pending operations visible to a scheduler, regardless of whether those operations are stored in a standalone queue object or inside the batching worker.

\begin{table}[t]
\centering
\small
\begin{tabular}{L{0.18\linewidth}L{0.35\linewidth}L{0.39\linewidth}}
\toprule
\textbf{Queue} & \textbf{Request and consumer} & \textbf{Scheduling objective} \\
\midrule
Training Queue & RL trajectories or SFT batches consumed by preprocessing workers and the Training Model Service. & Respect priority, readiness, in-flight limits, and policy staleness while retaining data until optimization completes. \\
Inference Queue & Observation batches from rollout or evaluation consumed by the Inference Model Service. & Minimize response latency while grouping requests that share a valid frozen-prefix computation. \\
\bottomrule
\end{tabular}
\caption{The two logical queue--scheduler pairs in \name{}. Each queue preserves tenant identity but optimizes a different execution objective.}
\label{tab:app-dual-queue}
\end{table}

The Training Scheduler and Inference Scheduler do not impose a global barrier on one another.
An RL tenant may continue submitting action requests to the Inference Queue while a previously collected trajectory waits in the Training Queue, and evaluation may use the Inference Queue without creating any optimization work.
The only cross-queue dependency is policy publication: after an RL update, the Training Model Service publishes the new tenant parameters to the Inference Model Service for subsequent requests.

\subsection{Training Job Descriptor and Data Binding}

RL and SFT workloads use the same unit of training scheduling: a training job descriptor maintained by the Training Scheduler and a data partition retained by the Training Queue.
Table~\ref{tab:app-job-descriptor} summarizes the fields that are relevant to orchestration.
The descriptor contains no training tensors; its \texttt{partition\_id} is the capability by which the producer, preprocessing stages, and actor refer to the corresponding payload.
The Training Scheduler creates this identifier from the training job, domain, tenant, and task identities, so concurrently active training jobs cannot accidentally address the same partition.

\begin{table}[t]
\centering
\small
\begin{tabular}{L{0.27\linewidth}L{0.65\linewidth}}
\toprule
\textbf{Field group} & \textbf{Role in execution} \\
\midrule
Identity & \texttt{job\_id}, \texttt{policy\_key}, \texttt{tenant\_id}, and \texttt{task\_type} identify the unit of work and its tenant-local policy. \\
Data binding & \texttt{partition\_id} binds control metadata to a Training Queue partition; \texttt{sample\_count} determines the planned amount of data. \\
Compatibility & \texttt{domain} selects the backend, while \texttt{schema\_signature} in training-job metadata identifies the canonical VLA interface. \\
Ordering & \texttt{priority}, \texttt{created\_seq}, and \texttt{ready\_seq} determine admission and lease order. \\
RL consistency & \texttt{behavior\_policy\_version} records the policy used for collection; \texttt{max\_policy\_staleness} bounds its distance from the actor. \\
Lifecycle & \texttt{producer\_role}, \texttt{consumer\_role}, \texttt{state}, timestamps, and terminal reasons support routing and monitoring. \\
\bottomrule
\end{tabular}
\caption{Control-plane fields of a VLA training-job descriptor. The schema signature is carried in the descriptor's metadata map.}
\label{tab:app-job-descriptor}
\end{table}

For each tenant--task policy, admission is controlled by a priority, a maximum number of in-flight training jobs, and an optional completion target.
A reservation is rejected when the in-flight or completion bound has already been reached; the producer then tries another eligible tenant or waits for progress.
Once training jobs become ready, the actor selects the highest-priority candidate and uses \texttt{ready\_seq} to preserve FIFO order among training jobs with equal priority.
Thus, \texttt{created\_seq} describes when work entered the system, whereas \texttt{ready\_seq} describes when all producer-side dependencies were satisfied.

Table~\ref{tab:app-job-states} gives the normal state sequences.
Actor-forward is optional for RL algorithms that do not require newly computed action log-probabilities or values.
The intermediate \texttt{computing\_*} states are leases held by the corresponding preprocessing service; they prevent two service replicas from processing the same stage concurrently.

\begin{table}[t]
\centering
\small
\begin{tabular}{L{0.17\linewidth}L{0.75\linewidth}}
\toprule
\textbf{Path} & \textbf{Normal state sequence} \\
\midrule
SFT & \texttt{reserved} $\rightarrow$ \texttt{ready} $\rightarrow$ \texttt{leased} $\rightarrow$ \texttt{completed} \\
RL & \texttt{reserved} $\rightarrow$ \texttt{computing\_advantages} $\rightarrow$ \texttt{ready} $\rightarrow$ \texttt{leased} $\rightarrow$ \texttt{completed} \\
Evaluation workload & Out-of-band inference and environment interaction; no training job or Training Queue partition is created. \\
\bottomrule
\end{tabular}
\caption{Normal training-job lifecycle for RL and SFT workloads. Brackets denote an optional RL stage. A nonterminal training job may instead enter \texttt{failed}; a stale RL training job enters \texttt{dropped}.}
\label{tab:app-job-states}
\end{table}

\subsection{Inference Queue Composition}

Each environment step creates an inference request containing the tenant identity, schema signature, observations, robot state, language instruction, and policy version.
The Inference Scheduler examines requests in arrival order and forms the largest FIFO-compatible prefix allowed by the batching timeout and maximum batch size.
Compatibility is determined at the shared frozen-prefix boundary rather than by tenant identity: accepted requests must use the same base-model interface and must satisfy the model-specific rank and shape constraints.
Canonicalization and permitted padding then produce one physical encoder batch.

\begingroup
\footnotesize
\begin{verbatim}
async def submit_action_request(session, observation):
    request = InferenceRequest(
        tenant_id=session.tenant_id,
        session_id=session.id,
        schema_signature=session.schema_signature,
        policy_version=session.policy_version,
        observation=observation)
    future = inference_queue.enqueue(request)
    return await future

async def inference_worker():
    requests = await inference_scheduler.collect_prefix(
        inference_queue,
        max_batch_size=encode_batch_size,
        timeout=encode_batch_timeout,
        compatible=shared_prefix_compatible)
    batch, slices = canonicalize_and_pad(requests)
    shared_features = inference_model.encode(batch)
    for request, feature in split(shared_features, slices):
        action = inference_model.decode(
            tenant_id=request.tenant_id,
            policy_version=request.policy_version,
            feature=feature)
        inference_queue.resolve(request, action)
\end{verbatim}
\endgroup

After the shared encoder forward, features are partitioned according to the recorded slices and decoded by the corresponding tenant action modules.
The result is resolved to the originating session, preserving the request--response semantics expected by the environment loop.
Unlike the Training Queue, the Inference Queue does not retain a partition after the response is delivered.
Its scheduler may therefore be implemented together with the encode-batching worker without changing the logical dual-queue interface.

\subsection{Producer-Side Composition}

The RL producer records the current policy version before environment interaction.
Its sample count is the product of the number of environments and the interaction horizon.
Multiple RL training jobs may be produced asynchronously, subject to both a global concurrency limit and a per-tenant limit.
Each completed rollout advances independently to its next state; it does not wait for a slower tenant in the same scheduling window.

\begingroup
\footnotesize
\begin{verbatim}
async def produce_rl(tenant):
    cfg = tenant_config(tenant.id)
    version = await rollout.current_policy_version(tenant.id)
    job = training_scheduler.reserve(
        policy_key=policy(tenant.id, "rl"),
        domain="vla", tenant_id=tenant.id, task_type="rl",
        sample_count=cfg.env.num_envs * cfg.horizon,
        behavior_policy_version=version,
        producer_role="rollout", consumer_role="actor",
        metadata={"schema_signature": tenant.schema_signature})
    if job is None:                 # admission limit reached
        return
    await rollout.generate(
        rollout_id=job["created_seq"], job_metadata=job)
    next_state = "needs_advantages"
    training_scheduler.mark(job["job_id"], next_state)

async def produce_sft(tenant):
    cfg = tenant_config(tenant.id)
    job = training_scheduler.reserve(
        policy_key=policy(tenant.id, "sft"),
        domain="vla", tenant_id=tenant.id, task_type="sft",
        sample_count=sft_num_samples(cfg),
        behavior_policy_version=None,
        producer_role="sft", consumer_role="actor",
        metadata={"schema_signature": tenant.schema_signature})
    if job is None:
        return
    try:
        batch = build_vla_sft_batch(
            cfg, rollout_id=job["created_seq"],
            num_samples=job["sample_count"])
        data = to_training_queue_data(batch)
        await training_queue.async_put(
            data=data, partition_id=job["partition_id"])
    except Exception as error:
        await training_queue.async_clear_partition(
            partition_id=job["partition_id"])
        training_scheduler.fail(job["job_id"], reason(error))
        raise
    training_scheduler.mark(job["job_id"], "ready")
\end{verbatim}
\endgroup

The two producers deliberately terminate at different states.
SFT examples already contain supervised targets and can be consumed immediately.
An RL partition, by contrast, may require actor-forward fields and must be augmented with advantages and returns before it is declared ready.
The rollout generator and the SFT batch builder both attach tenant identity and schema metadata to every model-facing sample; RL samples additionally carry the behavior-policy version.

\subsection{RL Preprocessing Stages}

The optional actor-forward stage reads the rollout partition, evaluates the behavior actions under the resident actor, and attaches the log-probability and value fields needed by the learning objective.
The advantage stage then obtains batches through the Training Queue's two-step metadata/data interface, computes algorithm-specific advantages and returns, and writes the derived fields back against the same sample metadata.
The concrete estimator may be PPO-, GRPO-, or return-based; it does not change the scheduling protocol.

\begingroup
\footnotesize
\begin{verbatim}
async def advantage_stage():
    job = training_scheduler.lease_stage(
        state="needs_advantages", consumer_role="actor",
        domain="vla", leased_state="computing_advantages")
    if job is None:
        return
    try:
        for batch_size in global_batch_plan(job):
            meta = await training_queue.async_get_meta(
                data_fields=advantage_input_fields(),
                batch_size=batch_size,
                partition_id=job["partition_id"],
                task_name="compute_advantages_and_returns")
            samples = await training_queue.async_get_data(meta)
            derived = compute_advantages_and_returns(samples)
            await training_queue.async_put(data=derived, metadata=meta)
        training_scheduler.mark(job["job_id"], "ready")
    except Exception as error:
        training_scheduler.fail(job["job_id"], reason(error))
        raise
\end{verbatim}
\endgroup

These stage leases apply only to control metadata.
The rollout samples remain in the original partition, and derived tensors are associated with the sample metadata rather than copied into the Training Scheduler.
Consequently, actor-forward and advantage workers can be independently deployed or omitted without changing the actor's final input contract.

\subsection{Actor Consumption and Version Consistency}

The actor is the common consumer for ready RL and SFT training jobs.
Before reading a leased RL partition, it compares the current tenant version $v_{\mathrm{actor}}$ with the recorded behavior version $v_{\mathrm{behavior}}$.
The training job is stale when
\begin{equation}
v_{\mathrm{actor}} - v_{\mathrm{behavior}}
> \Delta_{\max},
\end{equation}
where $\Delta_{\max}$ is the tenant policy's configured staleness bound.
This check is not applied to SFT because demonstrations are independent of the current policy.

\begingroup
\footnotesize
\begin{verbatim}
def consume_training_job(actor_versions):
    job = training_scheduler.lease_next(
        consumer_role="actor", domains=["vla"])
    if job is None:
        training_scheduler.wait_for_ready(
            consumer_role="actor", domains=["vla"], timeout_s=1)
        return

    partition = job["partition_id"]
    if (job["task_type"] == "rl" and
            job["max_policy_staleness"] is not None and
            job["behavior_policy_version"] is not None and
            actor_versions[job["tenant_id"]]
            - job["behavior_policy_version"]
            > job["max_policy_staleness"]):
        training_queue.async_clear_partition(partition)
        training_scheduler.drop(job["job_id"], "policy_stale")
        return

    try:
        metrics = actor.train_job(job)
        update_versions(actor_versions, metrics)
        maybe_checkpoint(job["tenant_id"], actor_versions, metrics)
        training_queue.async_clear_partition(partition)
        training_scheduler.complete(
            job["job_id"],
            metadata={"actor_versions": actor_versions})
    except Exception as error:
        training_queue.async_clear_partition(partition)
        training_scheduler.fail(job["job_id"], reason(error))
        raise
\end{verbatim}
\endgroup

Internally, \texttt{train\_job} derives a global batch plan from \texttt{sample\_count}, repeatedly obtains batch metadata and tensors from the leased partition, and validates their tenant and schema annotations.
For distributed training, each optimizer microbatch is required to contain one tenant--schema group.
The actor activates that tenant's action module and optimizer, computes the task-specific loss, performs backward and optimizer steps, increments only that tenant's policy version, and saves the updated tenant state.
RL updates subsequently publish the new tenant parameter payload to inference; SFT training jobs keep their result in actor and checkpoint state until a later workflow requests serving or export.
At no point are losses, gradients, optimizer states, or version counters averaged across tenants.

Partition cleanup occurs for all three terminal outcomes shown above.
A successfully consumed partition is cleared before the training job is marked \texttt{completed}; a stale partition is cleared before \texttt{dropped}; and producer or consumer exceptions mark the training job \texttt{failed} after best-effort cleanup.
Terminal training-job state is retained as Training Scheduler metadata, which permits progress and failure accounting after its tensors have been released.

\end{document}